\documentclass[]{pasj02} 
\usepackage[switch,mathlines]{lineno} 

\jyear{2025}
\Received{}
\Accepted{}

\begin{document} 

\title{
Far-infrared synchrotron properties of the inner lobes of the radio galaxy Centaurus A 
revealed with the Herschel observatory}

\author{
Naoki \textsc{Isobe},\altaffilmark{1,2}
 \email{n-isobe@ir.isas.jaxa,jp} 
Motoki \textsc{Kino},\altaffilmark{3,4}
    \orcid{0000-0002-2709-7338}
Takao \textsc{Nakagawa},\altaffilmark{5,1}
    \orcid{0000-0002-6660-9375}
Shunsuke \textsc{Baba},\altaffilmark{1}
    \orcid{0000-0002-9850-6290} 
Makoto \textsc{Tashiro}, \altaffilmark{6}
    \orcid{0000-0002-5097-1257}    
and
Hiroshi \textsc{Nagai} \altaffilmark{4,7}
    \orcid{0000-0003-0292-3645}
    }

\altaffiltext{1}{
    Institute of Space and Astronautical Science (ISAS), 
    Japan Aerospace Exploration Agency (JAXA), 
    3-1-1 Yoshinodai, Chuo-ku, Sagamihara, Kanagawa 252-5210, Japan}
\altaffiltext{2}{
    Department of Physics and Astronomy, 
    Kwansei Gakuin University, 1 Gakuen Uegahara, Sanda, Hyogo 669-1330, Japan}
\altaffiltext{3}{
    Division of Liberal Arts, 
    Kogakuin University of Technology \& Engineering
    2665-1 Nakano-machi, Hachioji, Tokyo 192-0015, Japan }
\altaffiltext{4}{
    National Astronomical Observatory of Japan,
    2-21-1 Osawa, Mitaka, Tokyo 181-8588, Japan}
\altaffiltext{5}{
    Advanced Research Laboratories, 
    Tokyo City University, 1-28-1 Tamazutsumi, Setagaya, Tokyo 158-8557, Japan}
\altaffiltext{6}{
    Department of Physics, Saitama University, 
    255 Shimo-Okubo, Sakura-ku, Saitama, Saitama 338-8570, Japan}
\altaffiltext{7}{
    Department of Astronomical Science, 
    The Graduate University for Advanced Studies, 
    SOKENDAI, 2-21-1 Osawa, Mitaka, Tokyo 181-8588, Japan}



\maketitle

\begin{abstract}
Diffuse far-infrared synchrotron emission 
filling the northern inner lobe of the radio galaxy Centaurus A 
is investigated with the Spectral and Photometric Imaging Receiver 
onboard the Herschel observatory at its three photometric bands. 
The far-infrared flux density spatially integrated over the lobe 
is measured as $S_{\rm \nu} = 1.63 \pm 0.05$ Jy 
at the wavelength of $500$ $\mu$m (the frequency of $600$ GHz). 
A comparison between the far-infrared spectral index derived with Herschel ($\alpha = 1.32 \pm 0.19$) 
and the radio index ($\alpha = 0.66 \pm 0.04$) 
suggests a spectral break between these frequency ranges. 
The change of the spectral index through the break is indicated 
to be consistent with that of the standard cooling break ($\Delta \alpha = 0.5$) 
predicted for particle acceleration under the continuous energy injection condition. 
A broken power-law model incorporating the standard cooling break 
yields the break frequency as $\nu_{\rm b} = 218 \pm 83$ GHz. 
From the measured cooling break frequency, the magnetic field of the northern inner lobe 
is evaluated as $B \gtrsim 100$ $\mu$G.
It is quantitatively estimated that 
the adiabatic cooling puts only a minor impact 
on the derived magnetic field.
This magnetic field is higher than that under the minimum-energy condition 
by more than a factor of $5$.
In addition, the derived magnetic field of the lobe is suggested to be  
at least by a factor of $4$ stronger than that of the inner-jet region 
implied in the previous very-high-energy gamma-ray study.
Even if the line-of-sight orientation of the lobe is considered 
in its possible extreme case, the magnetic field is found to be reduced only by a factor of 2, 
and the above arguments about the strong magnetic field basically holds.
The science impact of this result is discussed from the viewpoints of jet energetics, 
and of ultra-high energy cosmic rays. 
\end{abstract}


\section{Introduction}
\label{sec:intro}
Despite extensive research activities for more than several decades, 
the source of Ultra-High Energy Cosmic Rays (UHECRs) 
with an energy higher than $\sim10^{18}$ eV has remained 
one of the most important unsettled mysteries in modern astrophysics. 
Relativistic jets hosted by active galactic nuclei
and their related structures, including hot spots, knots and lobes,
are often discussed as a promising candidate for the UHECR accelerator 
(e.g., \cite{hillas1984,kotera2011}).  
Non-thermal synchrotron and inverse Compton (IC) radiations 
widely detected from these objects 
(e.g., \cite{meisenheimer1997,isobe2002,hardcastle2004,kataoka2005,uchiyama2006})
are regarded as concrete observational evidence 
of particle acceleration within them.
Actually, the UHECR anisotropy on a spatial scale of a few 10 degree 
suggested in recent observations \citep{aab2018,tkachev2021}
is frequently attempted to be associated with nearby radio galaxies 
\citep{matthews2018,abdul2024,mollerach2024}. 

It is commonly noticed that magnetic fields play 
a variety of fundamental roles within the particle acceleration process.  
As a first step to investigate the possible astrophysical UHECR sources, 
the relation between their magnetic field and size 
(the so-called Hillas diagram; \cite{hillas1984,kotera2011}) 
is widely utilized. 
Therefore, in order to figure out whether the jets and 
their associated structures really act as an UHECR acceleration site, 
a systematic measurement of their magnetic field is 
of significant importance.

\begin{figure*}[htp]
    \centering
    \includegraphics[width=0.95\linewidth]{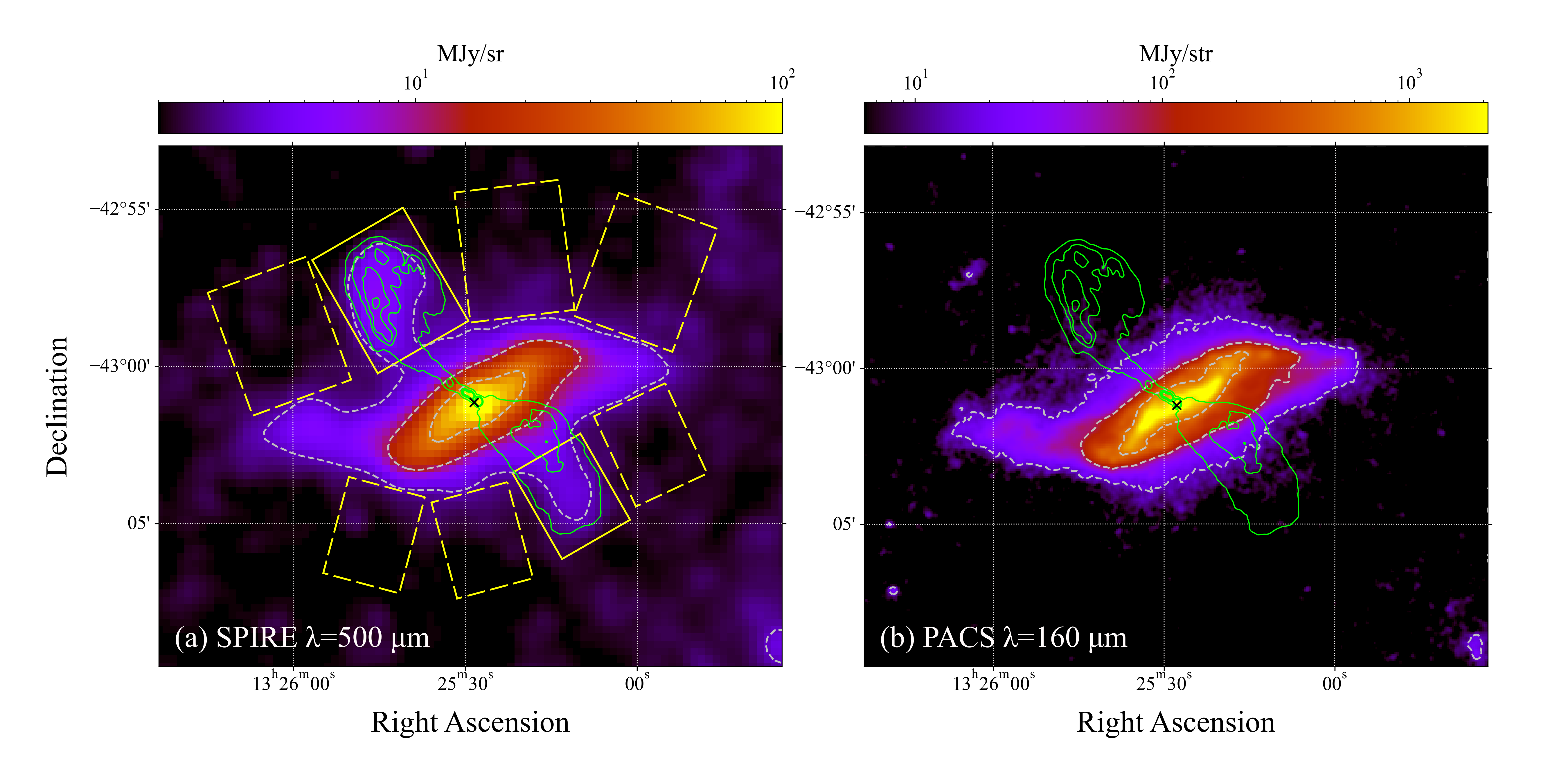}
    \caption{(a) SPIRE image at the wavelength of $\lambda=500$ $\mu$m 
        around the host galaxy and inner lobes of Centaurus A.
        The image is weakly smoothed with a two-dimensional Gaussian function 
        of which the radius is 1 pixel ($14$ arcsec).
        The short-dashed contours show the surface-brightness levels
        of $3.4$, $11.9$ and $42$ MJy str$^{-1}$.
        The 1.6 GHz VLA image \citep{hardcastle2006} is superposed 
        with the thin solid contours. 
        The cross near the peak of the FIR emission points the position of the nucleus \citep{hunt2021AJ}.
        The solid and long-dashed rectangles indicate 
        the photometric regions of the source 
        and fore/background FIR emission, respectively. 
        (b) PACS image at $\lambda=160$ $\mu$m, 
        weakly smoothed with the Gaussian function 
        with a radius of $3.2$ arcsec 
        (corresponding the pixel size of the PACS $160$ $\mu$m  image). 
        The contours are drawn at the levels of $15$, $86.6$ and $500$ MJy str$^{-1}$. 
        The J2000.0 coordinate system is adopted for the both panels.\\
        }
    \label{fig:image}
\end{figure*}

Since the advent of the Chandra X-ray Observatory, in particular, 
a flux ratio of the IC X-ray emission to radio synchrotron one
\citep{harris1979,band1985} has been a canonical tool 
to gauge the magnetic field in the radio-galaxy jets.
Thanks to relatively broad utilization of this method 
to the hot spots and lobes of Fanaroff-Riley type-II 
(FR-II; \cite{fanaroff1974}) radio galaxies,
the majority of them is suggested to exhibit a magnetic field 
nearly consistent with or much weaker than 
the equipartition and/or minimum-energy values
(e.g., \cite{isobe2002,hardcastle2004,kataoka2005,isobe2005}).
However, \citet{isobe2025} argued that 
the IC method is possibly more sensitive to 
objects with a higher electron dominance (and hence weaker magnetic field),
since a higher electron energy density tends 
to yield a higher IC X-ray intensity. 

The integrated synchrotron spectrum 
from the hot spots and lobes of FR-II radio galaxies 
is inferred to show a break feature 
in the range from the radio to far-infrared (FIR) 
or submillimeter frequencies. 
The break frequency is determined by the interplay 
between the adiabatic loss and radiative cooling \citep{inoue1996},
and hence, the feature is usually referred to as the cooling break. 
\citet{isobe2025} proposed the cooling break 
as another reliable estimator of the magnetic field
independently of the IC X-ray flux.
The method was pioneeringly launched out into a few hot spots,
by making use of FIR data obtained with the Herschel observatory
\citep{isobe2020,sunada2022,isobe2023}.
As a result, the mid-to-far infrared excess discovered  
from the western hot spot of the radio galaxy Pictor A 
is suggested to require a magnetic field stronger than 
that under the minimum energy condition 
by nearly an order of magnitude \citep{isobe2020,isobe2023}. 
However, the systematic application of the FIR data 
into the particle acceleration phenomena associated 
with the astrophysical jets has been still in its infancy.

The radio galaxy Centaurus A is widely recognized 
as hosting the nearest active galactic nucleus. 
Owing to its proximity, the object has been frequently featured 
as a promising candidate for the origin of the UHECRs 
\citep{romero1996,matthews2018,mollerach2022,mollerach2024}.
Usually Centaurus A is classified as an FR-I radio source. 
The radio images of this object (e.g., \cite{burns1983}) 
reveal its prominent inner jet structure 
emanating from the nucleus especially toward the northeast direction.
In addition, the radio galaxy is known to be accompanied 
with the inner lobes in an angular scale of $\sim 10'$ 
around the nucleus and inner jet.

The inner jet of Centaurus A has been extensively 
investigated in a variety of wavelength ranges 
(e.g., \cite{kraft2002,hardcastle2003,kataoka2006,hardcastle2006,
tingay2009,hess2020,janssen2021,bogensberger2024}).
In the X-ray band, the inner jet is reported to be decomposed 
into more than $30$ knots and a diffuse component 
\citep{kraft2002,kataoka2006}.
Their X-ray emission is naturally ascribed to the synchrotron radiation 
from highly relativistic electrons.
Centaurus A is also known 
as a Very-High-Energy (VHE) gamma-ray emitter \citep{aharonian2009}.
Recent VHE observations \citep{hess2020} indicate that 
the VHE emission originates in the central region of Centaurus A 
including the inner jet. 
By attributing the VHE gamma-ray emission to the IC radiation 
from the diffuse component in the inner jet,
the magnetic field is estimated as $B\sim 20$ $\mu$G \citep{hess2020}.
\citet{hardcastle2011} tried to reproduce simultaneously 
the spatial distribution of the synchrotron radio emission 
from the diffuse component 
and the spectral energy distribution 
in the radio to VHE-gamma-ray frequency ranges.
They created a sophisticated spectral calculation code, 
where all the possible IC seed photon fields are included 
and all the relevant physical processes are taken into account. 
As a result, it is suggested that the magnetic field of the inner jet 
is nearly consistent to or possibly stronger than the equipartition value.

Meanwhile, the interior of the inner lobes of Centaurus A 
has remained unexplored in the X-ray band, 
because of relatively severe contamination from other X-ray components 
including the thermal emission from the interstellar medium 
associated with the host galaxy \citep{kraft2003}.  
At the southwest periphery of the southern inner lobe, 
the Chandra data unveiled a shell-like X-ray enhancement \citep{kraft2003}.
The X-ray shell is interpreted as corresponding to a strong shock 
induced by the interaction between the interstellar medium 
and southern inner lobe \citep{kraft2007}, 
the latter of which is suggested to be supersonically expanding.
Through a detailed investigation into deep Chandra observations
\citep{croston2009},
the Mach number of the lobe expansion is estimated as $\sim8$,
and the X-ray emission from a large fraction of the shell 
is confirmed to be of synchrotron origin. 
Although non-thermal X-ray emission is also detected 
from the region on the southern inner lobe, 
this emission is concluded to be significantly dominated 
by the X-ray shell located in front of or behind the lobe.
In the case of the northern inner lobe,
no prominent X-ray shell surrounding it is found \citep{kraft2003, kraft2007}.
Instead, the eastern side of the northern inner lobe, 
which is also bright in the radio and mid-infrared ranges,  
is detected in the X-ray bands \citep{hardcastle2006}.
The multi-frequency spectrum from this area in the radio-to-X-ray range 
is naturally interpreted by the synchrotron radiation \citep{hardcastle2006}.

In summary, the IC technique has not yet been successfully applied 
to the magnetic-field evaluation of the inner lobes of Centaurus A. 
As a result,
previous multi-frequency spectral studies of the inner lobes 
still widely adopted the minimum-energy and/or equipartition magnetic field. 
\citep{clarke1992,brookes2006,hardcastle2006,weiss2008}. 
In contrast, \citet{isobe2025} predicted that 
the cooling break method definitely works on the inner lobes of Centaurs A,
when the FIR data are adequately appended. 
Actually, the FIR emission associated with the inner lobes 
was indicated in the FIR images obtained with the Herschel observatory
\citep{auld2012,parkin2012},
although their detailed spectral analysis has not yet been performed. 
Accordingly, the present paper aims at measuring the cooling break 
of the inner lobes of Centaurus A to evaluate their magnetic field 
by making most of the Herschel data.

Throughout the present paper, the distance to Centaurus A 
is assumed as $D=3.4$ Mpc \citep{israel1998},
because the value is considered to be 
most widely adopted to multi-frequency investigation 
into its inner jets and lobes 
(e.g., \cite{kraft2002,hardcastle2003,hardcastle2006,brookes2006}).
At this distance, the angular scale of $1''$ 
corresponds to the physical size of $16.5$ pc. 

\section{FIR data analysis}
\label{sec:ana}
\subsection{Herschel observations}
The inner region of the radio galaxy Centaurus A was previously observed 
twice with the photometer of the Spectral and Photometric Imaging Receiver 
(SPIRE; \cite{griffin2010}) 
aboard the Herschel observatory in the Large-Map mode
on 2009 December 27 (Obs. ID of 1342188663) and 2010 August 23 (Obs. ID of 1342203564).
The present study adopts the Level-2.5 science products,
generated by combining the data from these two observations.
From the Level-2.5 products, the SPIRE maps calibrated for diffuse sources
(i.e., labeled as {\tt extdPXW} with {\tt X} being {\tt L}, {\tt M} and {\tt S} 
for the wavelength of $\lambda = 500$, $350$, and $250$ $\mu$m) 
are mainly analyzed in the following. 

The FIR observation toward this object with the photometer installed 
on the Photodetector Array Camera and Spectrometer (PACS; \cite{poglitsch2010}) 
was conducted in the  Scan-map mode (Obs. IDs of 1342188855 and 1342188856)
on 2010 January 2.
The PACS blue channel was operated at the wavelength of $\lambda = 70$ $\mu$m.
The PACS Level-2.5 science data contain 
two types of image products applicable to diffuse sources 
(the Unimap and JScanam ones).
After checking that the results from these two image products 
are consistent with each other,
the analysis below employs the Unimap products. 

\subsection{FIR image} 
\label{sec:fir_image}
Figure \ref{fig:image}a displays the SPIRE image 
at the wavelength of $\lambda = 500$ $\mu$m 
(or the corresponding frequency of $\nu = 600$ GHz), 
while figure \ref{fig:image}b plots the PACS one 
at $\lambda = 160$ $\mu$m ($\nu = 1.87$ THz). 
The $1.6$ GHz radio image \citep{hardcastle2006},
obtained with the Very Large Array (VLA), 
is superposed on both maps with the thin solid contours. 
The brightest feature found on these FIR images is diffuse emission,
extending nearly in the east-west direction 
around the nucleus of Centaurus A (the cross in figure \ref{fig:image}).
It was reported that this component is naturally ascribed to cold dust
with a temperature of $T=20$--$30$ K,
associated with the disk of the host galaxy \citep{parkin2012}. 
A detailed investigation into the galactic dust emission 
is out of the scope of the present paper.

A comparison of the SPIRE with the radio images in figure \ref{fig:image}a
clearly reveals an additional faint diffuse FIR emission 
associated with  the inner lobes. 
At least at the wavelengths of $\lambda = 500$ and $350$ $\mu$m, 
the diffuse FIR emission seems to nearly entirely fill the lobes, 
with a relatively flat surface brightness without any prominent internal features.
The short-dashed contour shown in figure \ref{fig:image}a 
at the $500$ $\mu$m surface-brightness level of $3.4$ MJy str$^{-1}$
suggests that
a large part of the northern inner lobe is rather free from the dust FIR emission,
while nearly a half of the southern one is embedded in the galactic dust.
Figure \ref{fig:image}b indicates that 
the inner lobes have faded out at the PACS band.

\begin{figure}[tp]
    \centering
    \includegraphics[width=0.95\linewidth]{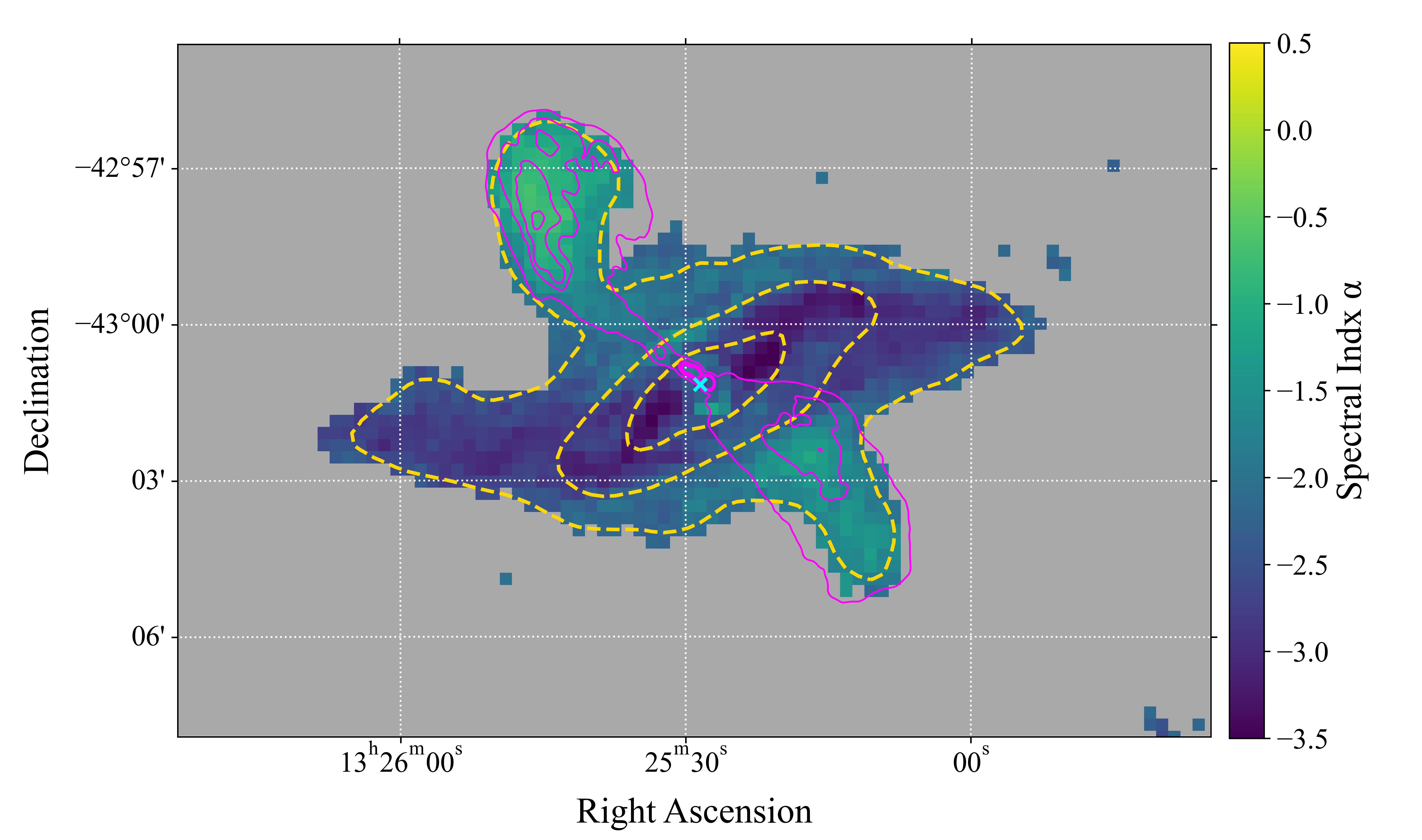}
    \caption{Two-point spectral-index map 
    between the wavelengths of $\lambda = 500$ $\mu$m and $350$ $\mu$m 
    (or the corresponding frequencies of $\nu = 600$ GHz and $857$ GHz,
    respectively).
    The region with a $500$ $\mu$m surface brightness of $< 3$ MJy str$^{-1}$
    is trimmed out and displayed in gray.
    The $500$ $\mu$m surface-brightness contours (the dashed lines) 
    and 1.6 GHz VLA contours (the thin solid lines) are taken 
    from figure \ref{fig:image}a.
    The nuclear position of Centaurus A \citep{hunt2021AJ}
    is shown with the cross. 
    The map is shown in the J2000.0 coordinates.\\
    }
    \label{fig:index_map}
\end{figure}

Figure \ref{fig:index_map} shows 
the spatial distribution of the spectral index, 
simply estimated by the ratio between the $500$ $\mu$m  
and $350$ $\mu$m ($\nu = 857$ GHz) images.
Throughout the present paper, 
the spectral index $\alpha$ is defined as $S_\nu \propto \nu^{-\alpha}$,
where $S_\nu$ denotes the flux density (e.g., in the Jy unit)
at the frequency $\nu$. 
The dust-dominated area on figure \ref{fig:image} exhibits
a "rising" spectrum with an index of $\alpha \lesssim -3$.
It is widely known that the dust spectrum is described with 
a modified black-body function, 
described as $S_\nu \propto \nu^\beta B_\nu(T)$,
where  $B_\nu(T)$ is the Plank function.
The index $\beta$ takes the dust emissivity function into account,
and its typical value is known as $\beta \sim 2$ (e.g., \cite{draine2003}).
In the Rayleigh–Jeans frequency regime,  
$B_\nu(T)$ is approximated as $\propto \nu^2$,
and thus, the spectral index of the dust emission is 
written as $\alpha = -\beta - 2$. 
It was reported that the dust spectrum in the host galaxy of Centaurus A 
is consistent with $\beta \simeq 2$ \citep{parkin2012}, 
and this dust emissivity index gives the spectral index of $\alpha \simeq -4$.
Therefore,
the spectral index within the dust-dominated region 
displayed in figure \ref{fig:index_map} 
is regarded as physically very reasonable. 

Figure \ref{fig:index_map} reveals that 
the inner lobes exhibit a significantly higher spectral index 
as $\alpha \gtrsim 0$.
This strongly indicates that 
the FIR emission from the inner lobes is not ascribed to the galactic dust. 
Instead, the FIR emission from the lobes 
is qualitatively thought to be of synchrotron origin,
based on its spectral index and its spatial coincidence to the radio emission.
Because the surface brightness of the FIR emission from the lobes 
is expected to be rather low, 
the index of the inner lobes presented in figure \ref{fig:index_map}
is probably biased to a lower value due to the contamination 
from the fore/background emission 
(of which the typical index is roughly estimated as $\alpha \sim -2.4$).
Therefore, the precise value of the FIR spectral index within the lobes 
should be reevaluated after the aperture photometry (see subsection \ref{sec:fir_phot}).

\subsection{FIR photometry} 
\label{sec:fir_phot}
\subsubsection{Northern inner lobe}
\label{sec:north_phot}

Aperture photometry is conducted to 
measure the FIR flux density of the northern inner lobe of Centaurus A.
The fore/background-inclusive FIR signal from this lobe is derived 
from the larger solid rectangle
drawn on the northern side of the galactic disk in figure \ref{fig:image}a. 
This source aperture,
which has a height and width of \timeform{250"} and \timeform{200"} 
respectively 
(the corresponding physical scale of $4.1$ and $3.3$ kpc 
at the distance of Centaurus A),
fully encompasses the northern inner lobe.
The fore/background FIR intensity is measured 
from the three long-dashed rectangles around the source aperture 
shown in figure \ref{fig:image}a.
The same size as of the source aperture is adopted 
for the fore/background apertures.
The FIR flux density averaged over the fore/background regions are 
subtracted from the flux density accumulated within the source region.
Following the standard procedure for the Herschel aperture photometry,
the standard deviation of the flux density 
evaluated from the fore/background apertures 
is regarded as the photometric error.
No color correction is applied,
because its impact is regarded as relatively small, 
i.e., $\sim 3$\% \citep{griffin2013} and $\sim 2$\% \citep{poglitsch2010}
for the SPIRE and PACS photometry, respectively, 
over the spectral-index range of $\alpha = 0$--$2$.

The result from the Herschel photometry on the northern inner lobe 
is tabulated in table \ref{tab:photometry}.
The FIR emission from the object is significantly detected 
at all the three SPIRE photometric bands. 
The $500$ $\mu$m flux density of the lobe is measured as 
$S_\nu (500~\mu{\rm m})= 1.63 \pm 0.05$ Jy. 
In contrast, the PACS photometer reveals no meaningful FIR signal from the lobe 
at both the red ($\lambda = 160$ $\mu$m) and blue ($\lambda = 70$ $\mu$m) channels.
This is consistent with the PACS image shown in figure \ref{fig:image}b.
From the PACS data, only a loose upper limit is derived 
as $S_\nu (160~\mu{\rm m}) \le 3.9$ Jy at the $3 \sigma$ confidence level,
as listed in table \ref{tab:photometry}.

\begin{figure}[t]
    \centering
    \includegraphics[width=0.95\linewidth]{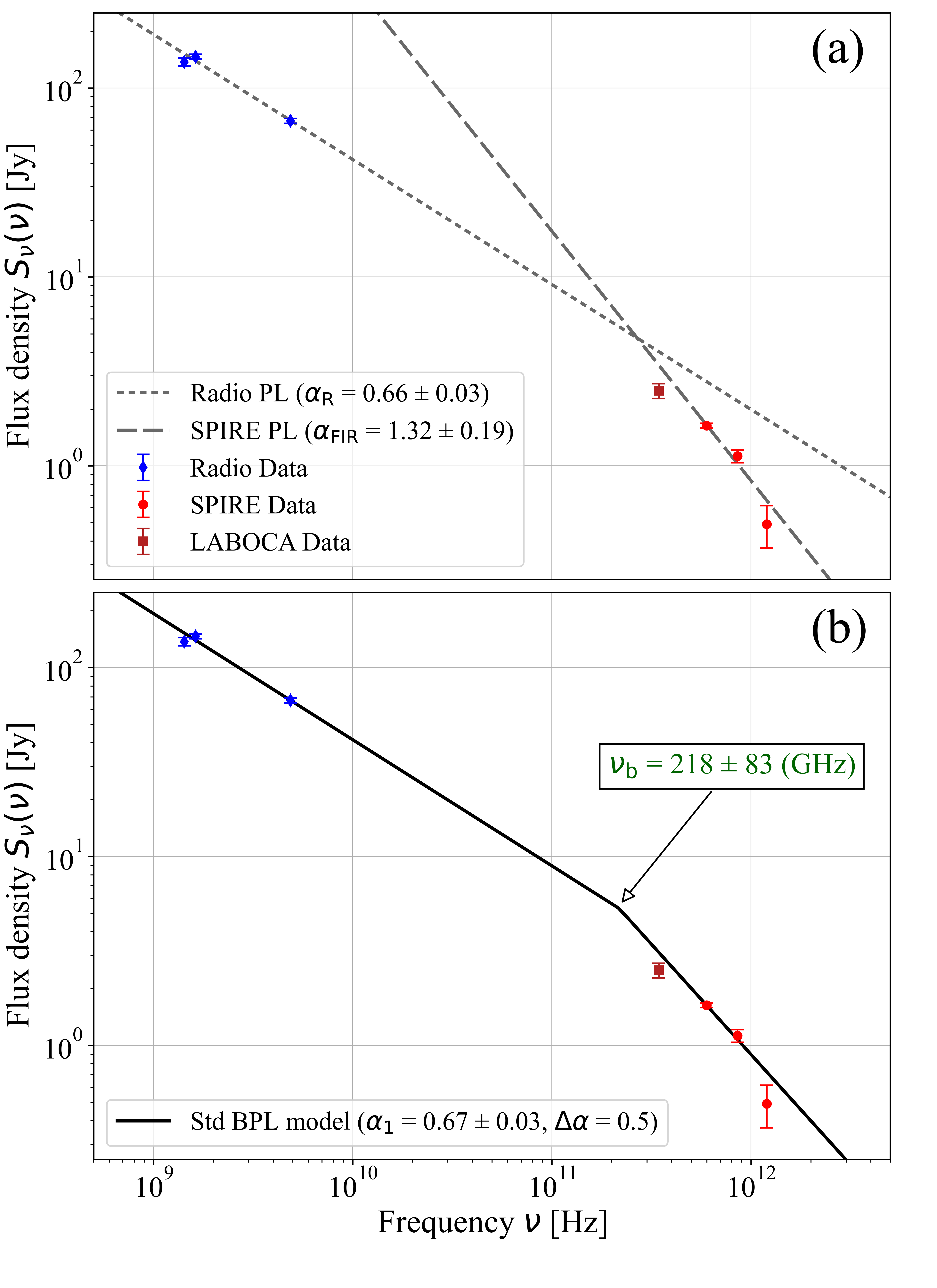}
    \caption{Synchrotron spectrum of the northern inner lobe of Centaurus A. 
    The  circles and  diamonds show the SPIRE and radio data, 
    respectively.
    The submillimeter flux density of the object, 
    obtained with LABOCA at the frequency of $\nu= 345$ GHz \citep{weiss2008}, 
    is plotted with the  square, 
    although this data point is omitted in the spectral fitting.
    In panel (a), the best-fit PL model to the SPIRE data derived 
    in sub-subsection \ref{sec:north_phot} is indicated with the dashed line, 
    while that to the radio data (see section \ref{sec:radio_phot}) 
    is drawn with the dotted line. 
    The solid line in panel (b) presents the Std BPL model
    (i.e., $\Delta \alpha = 0.5$ at the break),
    successfully describing the radio-to-FIR data.\\
    } 
    \label{fig:Sync_spec}
\end{figure}

The SPIRE spectrum of the northern inner lobe is 
displayed in figure \ref{fig:Sync_spec} with the circles.
As drawn with the dashed line in figure \ref{fig:Sync_spec}a,
the SPIRE data is approximated by a simple PL model.
The best-fit spectral index are evaluated as $\alpha_{\rm FIR} = 1.32 \pm 0.19$. 
The obtained index justifies that the color correction is
unimportant to the result of the SPIRE photometry 
(less than $1.5$ \% for this index value). 
Thanks to the fore/background subtraction,
the index of the lobe was improved from the value grasped from figure \ref{fig:index_map}, 
as mentioned in subsection \ref{sec:fir_image}.
Because this SPIRE spectrum is significantly softer than 
the typical radio spectrum of lobes of radio galaxies in the GHz range
($\alpha \gtrsim 0.5$; see section \ref{sec:radio_phot}), 
the SPIRE frequency range is suggested to reside in the cooling regime,
in the case of this object.

\begin{longtable}{lllll}
\caption{Summary of the FIR and radio photometry 
        for the northern and southern inner lobes.}
\label{tab:photometry}
\hline
Instrument  & $\lambda$ [$\mu$m] \footnotemark[$*$] & $\nu$ [GHz] \footnotemark[$*$]
            & \multicolumn{2}{c}{$S_\nu$ [Jy] \footnotemark[$\dag$]} \\
            &                    &       
            & Northern lobe      & Southern lobe  \\
\hline
\endfirsthead
\hline
Instrument  & $\lambda$ [$\mu$m] \footnotemark[$*$] & $\nu$ [GHz] \footnotemark[$*$]
            & \multicolumn{2}{c}{$S_\nu$ [Jy] \footnotemark[$\dag$]} \\
            &                    &       
            & Northern lobe      & Southern lobe  \\
\hline
\endhead
\hline 
\endfoot
\hline
\multicolumn{5}{l}{\footnotemark[$*$] 
    Reference wavelength and frequency for the photometry. } \\
\multicolumn{5}{l}{\footnotemark[$\dag$] 
    Fore/background-subtracted flux density. }\\
\multicolumn{5}{l}{\footnotemark[$\ddag$] 
    $3\sigma$ upper limit.} \\
\multicolumn{5}{l}{\footnotemark[$\S$] 
    The VLA images are taken from \citet{hardcastle2006}.}\\
\multicolumn{5}{l}{\footnotemark[$\|$] 
    The VLA image is taken from \citet{condon1996}.}
\endlastfoot
SPIRE   & $500$             & $600$  
        & $1.63 \pm 0.05$   &  $0.67 \pm 0.05$   \\
.....   & $350$             & $857$ 
        & $1.13 \pm 0.09$   &  $0.61 \pm 0.16$ \\
.....   & $250$             & $1.20\times10^3$   
        & $0.49\pm0.12$     & $\le 1.6$ \footnotemark[$\ddag$] \\
PACS    & $160$             & $1.87\times10^3$
        & $\le 3.9$ \footnotemark[$\ddag$]        
                            & $\le 3.8$ \footnotemark[$\ddag$] \\
\hline
VLA     & .....             & $4.87$ \footnotemark[$\S$] 
        & $ 67.1 \pm 2.0$   & $19.7 \pm 0.6$\\  
.....   & .....             & $1.63$ \footnotemark[$\S$]  
        &  $146.5 \pm 4.4$  & $55.4 \pm 1.7$\\  
.....   & .....             & $1.43$  \footnotemark[$\|$]  
        &  $137.4 \pm 6.9$  & $55.1 \pm 2.8$\\  
\end{longtable}

\subsubsection{Southern inner lobe}
\label{sec:south_phot}
The FIR intensity of the southern inner lobe is evaluated
in the manner similar to that adopted for the northern one 
(see sub-subsection \ref{sec:north_phot}).
The smaller solid and dashed rectangles 
illustrated on the southern side of the galactic emission in figure \ref{fig:image}a
represent the source and fore/background apertures,
respectively,  for the southern inner lobe.
These regions have a size of $190 \timeform{"} \times 150\timeform{"}$
($3.1$ kpc $\times 2.5$ kpc). 
In order to avoid the contamination from the dust FIR emission, 
the source region only covers 
nearly the southern half of the southern inner lobe.

Table \ref{tab:photometry} also charts
the SPIRE and PACS flux density of the southern inner lobe.
The lobe is found to exhibit significant FIR signals 
only at the two longer-wavelength SPIRE photometric bands, 
i.e., at $\lambda = 500$ and $350$ $\mu$m.
The $500$ $\mu$m flux density of the object is evaluated 
as $S_\nu(500~\mu{\rm m}) = 0.67 \pm 0.05$ Jy. 
At $250$ and $160$ $\mu$m, respectively, 
the SPIRE and PACS yield only a loose upper limit on the FIR flux density. 
The $500$--$350$ $\mu$m two-point spectral index 
of the southern inner lobe is given as $\alpha = 0.26 \pm 0.75$. 
The accuracy of this index is thought to be insufficient 
to specify whether or not the radiative cooling has 
a dominant impact in the FIR spectral range. 
Therefore, 
a further discussion on the properties of the southern inner lobe 
has been concluded to be out of the scope of the present paper.

\section{Radio photometry} 
\label{sec:radio_phot}
Radio photometry is performed on the northern inner lobe
to evaluate its low-frequency synchrotron spectrum. 
For this purpose, the $1.4$ GHz VLA image is taken from \citet{condon1996},
since the image has been widely utilized 
for multi-wavelength studies of the inner region of Centaurus A 
(e.g., \cite{brookes2006,weiss2008,parkin2012,hess2020,deOliveira2025}).
The present study additionally adopts the $1.6$ GHz and $4.9$ GHz VLA images 
picked up from \citet{hardcastle2006}.
Table \ref{tab:photometry} lists 
the radio flux density obtained from the source aperture 
shown in figure \ref{fig:image}a.
The photometric error of $3$\% is employed 
for the $1.6$ GHz and $4.9$ GHz VLA data after \citet{hardcastle2006}, 
whereas the $5$\% error is applied to the $1.4$ GHz result
as a conservative VLA photometric one \citep{perley2017}.
Only for a reference purpose, 
table \ref{tab:photometry} also tabulates 
the radio flux density integrated 
over the photometric aperture of the southern inner lobe.

The radio spectrum of the northern inner lobe, thus obtained, 
is plotted with the diamonds in figure \ref{fig:Sync_spec}.
The flux density of the lobe is measured as 
$S_\nu = 146.5\pm4.4$ Jy and $67.1\pm2.0$ Jy, respectively,
at the frequencies of $\nu=1.63$ GHz and $4.87$ GHz.
From the three VLA data points, 
the radio spectral index is measured as $\alpha_{\rm R} = 0.66 \pm 0.03$,
as shown with the dotted line in figure \ref{fig:Sync_spec}a. 
The obtained index is consistent with the picture that 
the impact of the radiative cooling is still negligible 
at the radio band.

\section{Multi-frequency synchrotron spectrum}
\label{sec:sync_spec} 
Figure \ref{fig:Sync_spec} compares 
the FIR synchrotron spectrum of the northern inner lobe of Centaurus A
obtained with the Herschel SPIRE photometer (the circle)
with the radio one (the diamonds).
In the figure, the total submillimeter flux of this lobe 
at the frequency of $\nu = 345$ GHz,
which is derived with the Large Apex Bolometer Camera (LABOCA) 
operated at the Atacama Pathfinder Experiment telescope \citep{weiss2008}, 
is shown with the square.
The LABOCA data point is not utilized in the spectral modeling below
because its photometric aperture is unclear. 
However, it is basically regarded as consistent 
with the spectral interpretation constructed by the radio and FIR data.

Figure \ref{fig:Sync_spec} suggests that 
the synchrotron spectrum of the object is significantly steeper 
in the FIR range than in the radio one.
The dashed and dotted lines in figure \ref{fig:Sync_spec}a show the 
PL models best-fit to the SPIRE and radio data, respectively. 
The spectral index of the SPIRE data ($\alpha_{\rm FIR} = 1.32 \pm 0.19$) is 
higher than that of the radio one ($\alpha_{\rm R} = 0.66 \pm 0.03$). 
These two PL models are found to intersect with each other
around the frequency of $\nu = 200$--$300$ GHz. 
It is indicated that the radio-to-FIR synchrotron spectrum of the lobe 
is possible to be approximated by a broken PL (BPL) model 
with a convex form on the $S_\nu$--$\nu$ plot. 

In order to examine particle acceleration phenomena 
associated with astrophysical jets emanating from active galactic nuclei,
the diffusive shock acceleration \citep{Begelman1984,Bell1978}
is widely accepted as the canonical mechanism.
This acceleration process is expected to create a synchrotron spectrum
with an index of $\alpha \gtrsim 0.5$.
Under the continuous energy injection condition \citep{Carilli1991}, 
which is employed as another standard premise, 
the synchrotron spectral index is inferred to change 
at the cooling break frequency by $\Delta \alpha = 0.5$ 
(the standard cooling break). 
The measured radio index of the northern inner lobe 
($\alpha_{\rm R} = 0.66 \pm 0.03$)
is regarded as reasonable from the diffusive shock acceleration.
The difference between the FIR and radio indices 
($\alpha_{\rm FIR}-\alpha_{\rm R} = 0.66 \pm 0.19$)
agrees with the standard cooling break within the errors. 
Thus, the radio-to-FIR synchrotron spectrum of the lobe is revealed   
to fully follow the canonical particle acceleration scenario.

Based on the above arguments, the synchrotron spectrum of the lobe
is described with a BPL model in which the standard cooling break 
with $\Delta\alpha = 0.5$ is incorporated (hereafter referred to as the Std BPL model).
The solid line in figure \ref{fig:Sync_spec}b shows the derived model curve.
The low-frequency spectral index of the Std BPL model ($\alpha_1 = 0.67 \pm 0.03$)
is confirmed to stay unchanged from the radio index. 
The high-frequency index, $\alpha_2 = \alpha_1 + 0.5 = 1.17 \pm 0.03$, 
is found to be within the error of the FIR index. 
As indicated with the arrow in figure \ref{fig:Sync_spec}b,
the break frequency is evaluated as $\nu_{\rm b} = 218 \pm 83$ GHz.
This frequency is reasonable from the comparison 
between the radio and FIR PL models shown in figure  \ref{fig:Sync_spec}a.

It is reported that the northern inner lobe of Centaurus A
is detected in the mid-infrared range with the Spitzer observatory
\citep{brookes2006,hardcastle2006}. 
\citet{hardcastle2006} identified three regions  
(the inner, middle and outer ones defined in their paper)
in the eastern side of the lobe,
which are bright in the mid-infrared frequency range. 
They derived the radio, mid-infrared, ultraviolet and X-ray flux densities 
of these regions. 
In addition, 
\citet{weiss2008} conducted 345 GHz submillimeter aperture photometry 
with the LABOCA instrument on the same regions. 
Among these regions, the middle and outer ones are found to be fully contained 
within the photometric area of the present study,
shown in figure \ref{fig:image}a. 

\begin{figure}[t]
    \centering
    \includegraphics[width=0.95\linewidth]{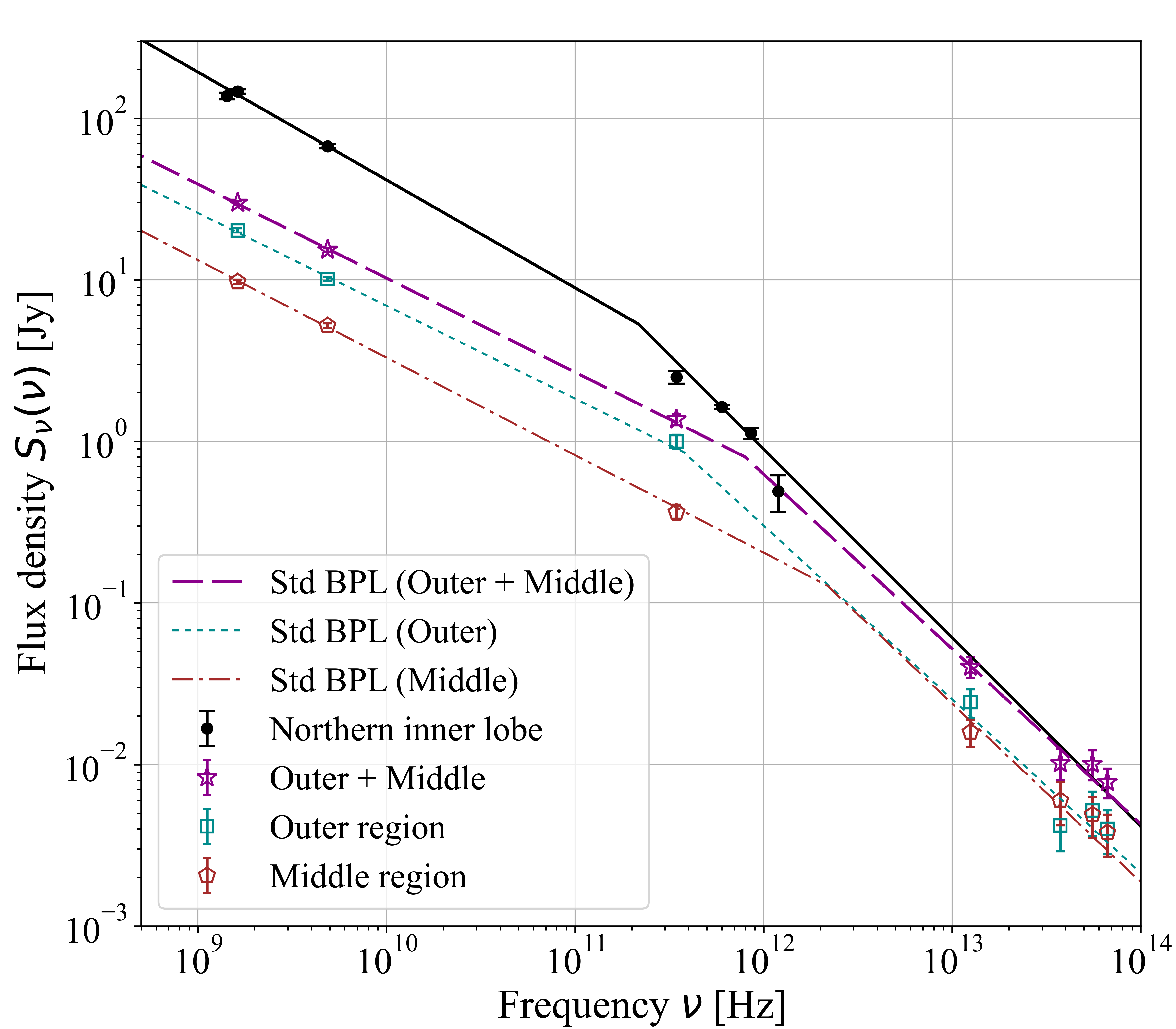}
    \caption{
    Comparison between the synchrotron spectrum of the northern inner lobe
    and those of the middle and outer regions inside the lobe
    (the pentagons and squares) defined in \citet{hardcastle2006}.    
    The sum spectrum of the middle and outer regions 
    is also plotted with the stars.
    The radio-to-FIR data of the northern inner lobe (the filled circles)
    and their best-fit Std BPL model (the solid line) 
    are same as those displayed in figure \ref{fig:Sync_spec}.
    The radio and mid-infrared data of the middle and outer regions 
    are taken from \citet{hardcastle2006}, 
    while their submillimeter data are picked up from \citet{weiss2008}.
    The dash-dotted and dotted lines show the Std BPL model ($\Delta \alpha = 0.5$) 
    best-fit to the spectra of the middle and outer regions,
    while the dashed line indicates the Std BPL model 
    approximating the sum spectrum of the middle and outer regions. \\
    }
    \label{fig:Comp_H06}
\end{figure}

In figure \ref{fig:Comp_H06}, 
the radio, submillimeter and mid-infrared flux densities 
of the middle and outer regions (the open pentagons and squares, respectively)
are compared with the radio-to-FIR ones of the entire northern inner lobe
taken from figure \ref{fig:Sync_spec}.
\citet{weiss2008} reported that 
the multi-frequency synchrotron spectrum in the radio to X-ray ranges
of these two regions is reproduced by the Std BPL model ($\Delta \alpha = 0.5$).
Thus, the radio-to-mid-infrared spectra of the middle and outer regions, 
and their sum (presented with the open stars in figure \ref{fig:Comp_H06}), 
are re-analyzed by using the Std BPL model, 
as drawn with the dash-dotted, dotted and dashed lines. 
The spectral index and the break frequency derived from the Std BPL model
are thought to be compatible to the results presented in \citet{weiss2008}.

It is revealed that the combination of the middle and outer regions 
only accounts for 58\% and 66\% of the total FIR flux density 
of the northern inner lobe at the wavelength of $\lambda = 500$ and $350$ $\mu$m 
($\nu = 600$ and $857$ GHz), respectively.
Figure \ref{fig:Comp_H06} suggests that the contribution 
from the middle and outer regions 
becomes lower toward the lower frequency range. 
The remaining FIR emission is attributed to more diffuse component 
extending over the whole lobe, including its western side. 
In other words, 
the cooling break frequency obtained for 
the spectrum integrated over the lobe is considered 
to reflect the condition in the lobe-scale diffuse component 
rather than that in the localized structures bright in the mid-infrared band.
In contrast, figure \ref{fig:Comp_H06} suggests that 
the high-frequency extrapolation of the Std BPL model of the northern inner lobe 
(the solid line) agrees with the combined spectrum of the middle and outer regions
(the open stars and the dashed line) in the mid-infrared range.
Therefore, the lobe-scale diffuse component is concluded to exhibit only 
a minor contribution above the FIR frequency range. 

\section{Magnetic-field evaluation}
\label{sec:B-field}
\subsection{Baseline method}
\label{sec:baseline}

Because the comparison between the radio and FIR synchrotron spectra
has successfully yielded 
the cooling break frequency of the northern inner lobe,
its magnetic field is evaluated after \citet{isobe2025}.
As far as the synchrotron cooling dominates the IC one,
as is the case for typical hot spots \citep{isobe2025},  
the magnetic field strength is derived 
from the cooling break frequency $\nu_{\rm b}$ as 
\begin{equation}
    B = \Big\{
    \frac{27 \pi e m_{\rm e} v^2 c}
        {\sigma_{\rm T}^2} L_{\rm c}^{-2} \nu_{\rm b}^{-1}  \Big\}^{\frac{1}{3}},  
\label{eq:B_from_break_freq}
\end{equation} 
where $e$, $m_{\rm e}$, $v$, $c$, $\sigma_{\rm T}$ and $L_{\rm c}$,
respectively, denote the elementary charge, electron mass, 
flow speed in the shock downstream evaluated in the shock frame,
speed of light, 
Thomson cross section,
and cooling length along the plasma flow. 
The method was actually applied to a few hot spots
for which the FIR spectrum was successfully measured 
\citep{sunada2022,isobe2020,isobe2023}.
\citet{isobe2025} proposed that the inner lobes of Centaurus A 
are one of the next targets to which the method is effectively applicable.

\begin{figure}[t]
    \centering
    \includegraphics[width=0.95\linewidth]{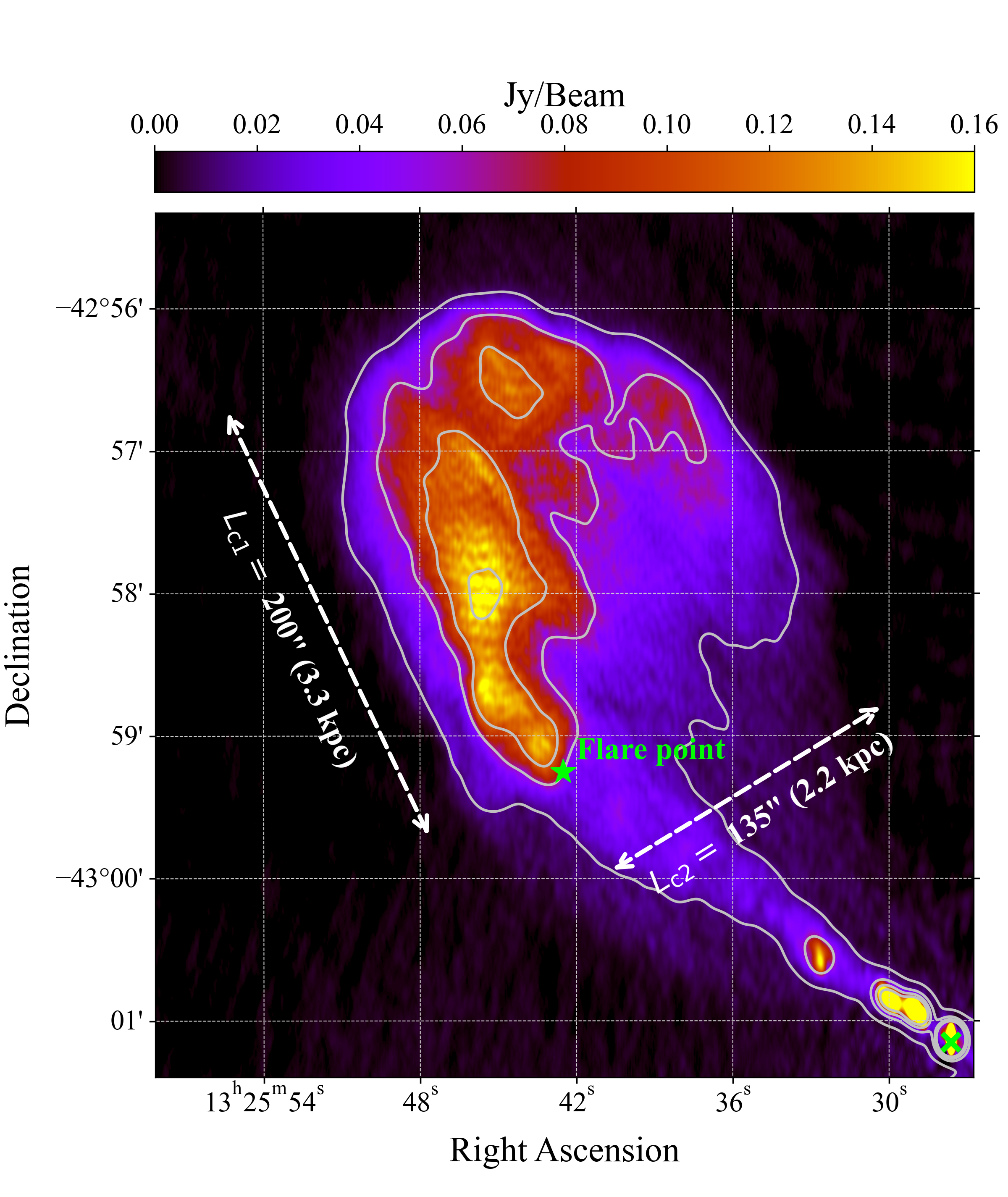}
    \caption{
    1.6 GHz VLA image of the northern inner lobe \citep{hardcastle2006},
    on which the cooling lengths adopted for the magnetic-field estimation
    are displayed with the dashed arrows denoted 
    as $L_{\rm c1}$ ($200''$ or $3.3$ kpc) and 
    $L_{\rm c2}$ ($135''$ or $2.2$ kpc).
    The star shows the flare point, 
    which is regarded as the energy injection region 
    from the jet to lobe \citep{hardcastle2006}.
    The cross at the bottom-right corner indicates 
    the position of the nucleus \citep{hunt2021AJ}. \\
    }
    \label{fig:Lcool}
\end{figure}

On the 1.6 GHz VLA image of the northern inner lobe of Centaurus A
\citep{hardcastle2006} shown in figure \ref{fig:Lcool},
the cooling lengths adopted for the magnetic-field evaluation 
are superposed with the dashed arrows 
indicated as $L_{\rm c1}$ and $L_{\rm c2}$.
\citet{hardcastle2006} reported that the area 
brightest in the mid-infrared band is located 
$\sim 3.4'$ away from the nucleus
as displayed with the filled star in figure \ref{fig:Lcool}.
They refereed to this area as the ``flare point''. 
The radio surface brightness is found to be drastically enhanced 
through the flare point toward the jet downstream.
Although diffuse X-ray emission was detected around the point
with a relatively high intensity,
the area behind it was claimed to exhibit no significant X-ray signals.
Based on these observational characteristics, 
\citet{hardcastle2006} concluded that the flare point represents 
the final spot of particle acceleration along the jet,
at which shock-energized high-energy particles 
are injected into the lobe. 
Therefore, the cooling length is evaluated by following this scenario. 

As mentioned in section \ref{sec:sync_spec},
the cooling break derived in figure \ref{fig:Sync_spec} 
is strongly suggested to reflect 
the physical condition of the lobe-scale diffuse emission.
Thus, the distance between the flare point and the edge of the lobe 
is reasonably considered as the cooling length. 
As an initial condition, 
the major and minor axes of the lobe are supposed 
to be simply aligned to the celestial plane, and thus, 
their angle to the line of sight is set at $\theta = 90^\circ$.
Here, two representative cases are considered for the cooling length.
The first case assumes that the cooling is dominated by the flow 
along the downstream jet.
The dashed arrow denoted as $L_{\rm c1}$ in figure \ref{fig:Lcool} 
shows the cooling length for this case.
The length of this arrow is estimated as $L_{\rm c1} = 200''$  
on the VLA image, which corresponds to the physical scale of $3.3$ kpc
at the distance of Centaurus A ($D=3.4$ Mpc; \cite{israel1998}). 
In contrast for the alternative case, 
the transverse flow is presumed to control the cooling. 
Thus, in this case, 
the cooling length is evaluated as $L_{\rm c2} = 135''$  
(or $2.2$ kpc in the physical size),
as also presented in figure \ref{fig:Lcool}.

\begin{figure*}[t]
    \centering        
    \includegraphics[width=0.95\linewidth]{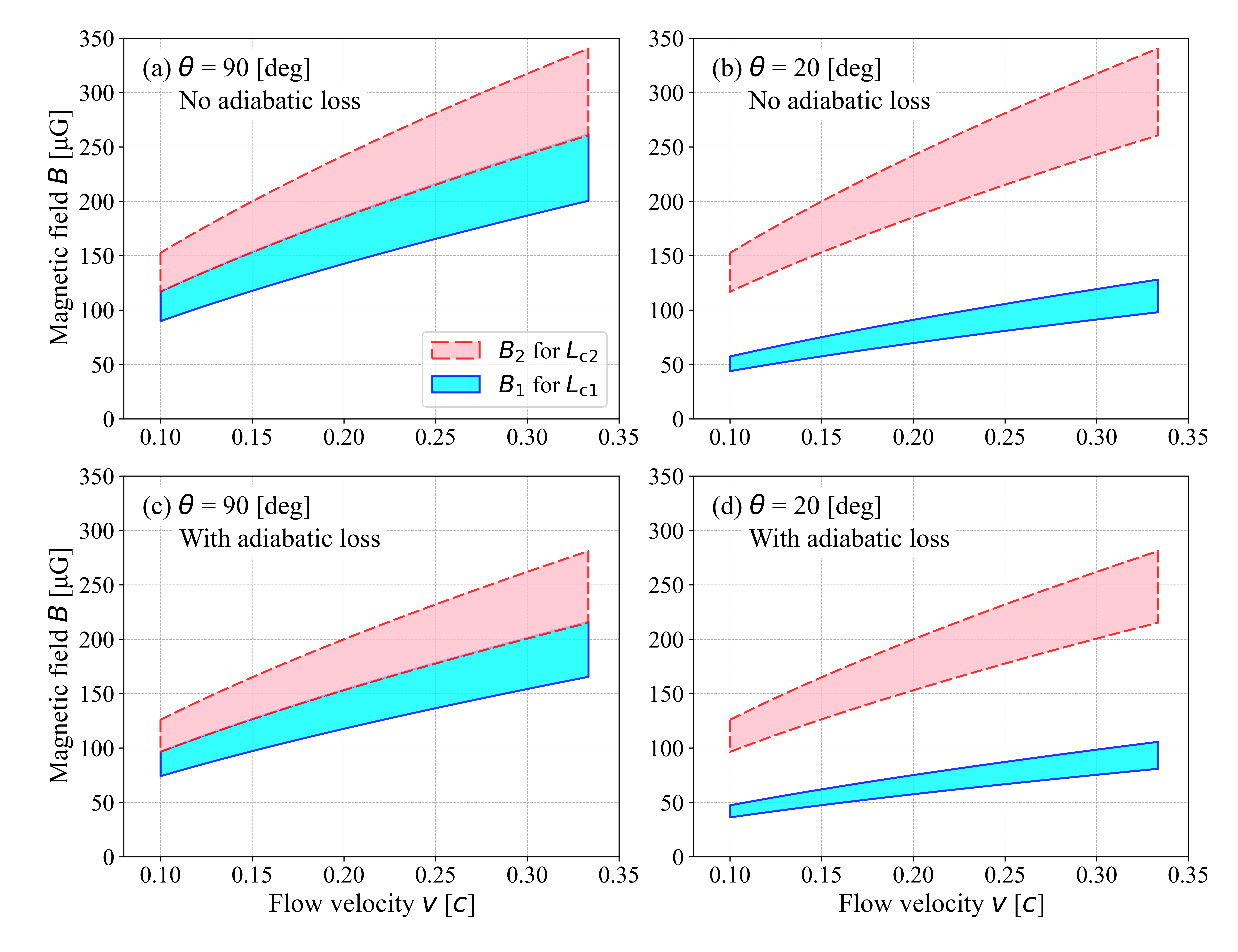}
    \caption{Magnetic field of the northern inner lobe 
    estimated from its cooling break ($\nu_{\rm b} = 218 \pm 83$ GHz),
    plotted against the downstream flow velocity $v$.
    The solid and dashed lines indicate the derived magnetic field,
    $B_1$ and $B_2$, respectively
    for the cooling length of $L_{\rm c1}$ and $L_{\rm c2}$.    
    The results in panels (a) and (b) are derived simply 
    by equation (\ref{eq:B_from_break_freq}) taken from \citet{isobe2025}, 
    where the adiabatic cooling is neglected.
    The results with the adiabatic loss included are presented in panels (c) and (d).
    In panels (a) and (c),  the major and minor axes of the lobe are assumed 
    to be parallel to the sky plane with their angle to the line of sight 
    being $\theta = 90^\circ$.
    Panels (b) and (d) employs $\theta=20^\circ$
    as the line-of-sight angle of the lobe major axis
    (i.e., the lower-end value suggested in \cite{hardcastle2003}).\\
    }   
    \label{fig:Mag}
\end{figure*}

Because of its low IC X-ray flux,
which is suggested to be significantly 
below a few 10 nJy at 1 keV \citep{hardcastle2006},
the radiative cooling in the northern inner lobe of Centaurus A
is considered as dominated by the synchrotron emission.
Therefore, 
the magnetic field evaluation via equation (\ref{eq:B_from_break_freq}) 
from the cooling break frequency ($\nu_{\rm b} = 218 \pm 83$ GHz)
is simply applicable to this object. 
The regions encompassed 
by the solid and dashed lines in figure \ref{fig:Mag}a
displays the derived magnetic field 
for the cooling length of $L_{\rm c1}$ and $L_{\rm c2}$, respectively,
as a function of the the downstream flow velocity $v$.

The upper limit on the flow velocity, $v=\frac{1}{3} c$, 
stems from the downstream flow velocity for
the highly relativistic strong shock \citep{Kirk1999,Kino2004}.
This flow velocity has been consistently employed 
in the previous studies for hot spots 
\citep{isobe2017,isobe2020,sunada2022,isobe2023,isobe2025}.
Combining the observed sub-relativistic apparent motion 
of inner knots located at the distance of a few $100$ pc 
(or $\gtrsim 10''$) from the nucleus of Centaurus A 
with a jet-counterjet comparison, 
\citet{hardcastle2003} inferred 
the intrinsic velocity of the inner jet as $(0.4$--$0.6) c$. 
By assuming the standard compression ratio 
of non-relativistic strong shocks ($r=4$),
the lower constraint on the downstream flow velocity 
is given as $v\gtrsim 0.1 c$. 
Thus, as shown in figure \ref{fig:Mag}a, 
the magnetic field in the northern inner lobe of Centaurus A 
is conservatively estimated as 
$B_1 \simeq 90$--$260$ $\mu$G for $L_{\rm c1}$ and
$B_2 \simeq 120$--$350$ $\mu$G for $L_{\rm c2}$.

\citet{hardcastle2003} implied that 
the inner jet located at the distance of $\sim 100$ pc 
from the nucleus of Centaurus A is possible 
to exhibit a relatively small angle to the line of sight 
as $\theta = 20^\circ$--$50^\circ$.
If the northern inner lobe is aligned to the inner jet, 
the longitudinal cooling length ($L_{\rm c1}$) shown above 
is thought to be considerably underestimated,
while in contrast, this jet orientation has basically 
no significant impact on the lateral cooling length ($L_{\rm c2}$).
As an extreme case, the magnetic field estimated 
for $\theta = 20^\circ$ is displayed in figure \ref{fig:Mag}b.
Because the longitudinal cooling length is corrected as $L_{\rm c1}/\sin20^\circ$
the magnetic field for $L_{\rm c1}$ is 
scaled by $(\sin 20^\circ)^{2/3} \simeq 0.5$.   
As a result, the magnetic field nearly halves as $B_1 \simeq 45 - 130$ $\mu$G.

\subsection{Impact of adiabatic cooling}
\label{sec:adiabatic}

The method presented in \citet{isobe2025} simply supposes that 
relativistic electrons lose their energy only via the radiative cooling 
(especially the synchrotron one in this case). 
However, in the case of the northern inner lobe of Centaurus A,
it is possible for the adiabatic cooling to put some notable impacts 
on the cooling break frequency,
because the lobe is expected to expand rapidly 
through ambient interstellar medium \citep{kraft2003,kraft2007}. 
Thus, in the following,
a brief evaluation is described 
on the effect of the adiabatic cooling onto the magnetic-field estimation.

For simplicity, it is presumed that the northern inner lobe is isobaric 
and stays in pressure balance with the surrounding interstellar medium. 
Under the condition, it is estimated that 
the fraction $\eta_{\rm ad}=(\kappa - 1)/\kappa$ 
of the initial energy supplied to the lobe from jet through the flare point 
is consumed via the adiabatic cooling,
where $\kappa$ denotes the adiabatic index (or the specific heat capacity ratio). 
The remaining fraction $\eta_{\rm c} = 1 - \eta_{\rm ad}=1/\kappa$
is thought to be eventually lost by the radiative cooling. 
These energy partition fractions have been often adopted 
to estimate the jet power from the pressure of the ambient matter 
\citep{allen2006,ito2008}. 
By adding the adiabatic loss to the radiative one,
the electron cooling rate is effectively enhanced
by a factor of $\eta_{\rm c}^{-1}$,
and correspondingly, 
the electron cooling timescale is reduced by a factor of $\eta_{\rm c}$.
As a result,
the cooling break frequency is scaled as $\eta_{\rm c}^2$.
Finally, after the adiabatic cooling is taken into account,
the magnetic field simply derived from equation (\ref{eq:B_from_break_freq})
is concluded to be modified by a factor of $\eta_{\rm c}^{\frac{2}{3}}$.

Figures \ref{fig:Mag}c and \ref{fig:Mag}d display 
the magnetic field of the northern inner lobe of Centaurus A,
in which the impact of the adiabatic cooling is introduced. 
Here, the adiabatic index of the relativistic plasma, 
$\kappa = 4/3$, is adopted,
and hence the fraction of the radiative cooling is given as $\eta_{c}=3/4$. 
For example, 
the magnetic field corresponding to the longitudinal cooling length ($L_{\rm c1}$) 
is found to be slightly weakened to $B_{\rm 1} \simeq 75$--$215$ $\mu$G 
for the lobe angle of $\theta=90^\circ$.
It is confirmed that 
the correction factor to account for the adiabatic loss is rather small 
as $\eta_{c} ^ {\frac{2}{3}} \simeq 0.83$.

\section{Discussion} 
\label{sec:discussion}
Based on the successful measurement of the cooling break frequency through the FIR data,
the magnetic field strength in the northern inner lobe of Centaurus A 
has been  estimated, for the first time without assuming 
the minimum-energy condition or equipartition one. 
In the following, 
the science impact of the result is briefly discussed,
by employing the $B_1$ value as the baseline.
The magnetic field from equation (\ref{eq:B_from_break_freq})
presented in subsection \ref{sec:baseline} is utilized below. 
Based on the consideration in subsection \ref{sec:adiabatic}, 
the scientific discussion is basically unaffected 
even by taking the adiabatic loss into consideration. 

\subsection{Magnetic-field dominance}
\label{sec:Bme}
Firstly, the derived magnetic field is compared with the minimum-energy value,
because this comparison is widely adopted as the standard procedure 
to investigate the particle-acceleration condition within the jets 
\citep{hardcastle2004,kataoka2005}.
Here, the minimum-energy field $B_{\rm me}$ is calculated 
after \citet{miley1980}.
The shape of the lobe is simply approximated by an ellipsoid 
with its major axis being the symmetric one. 
By referring to the radio image shown in figure \ref{fig:Lcool},
its major and minor diameter is roughly evaluated as 
$200''$  ($3.3$ kpc in the physical size) and 
$180''$  ($3.0$ kpc), respectively. 
The spectral index and reference flux density 
are taken from the best-fit Std BPL model 
shown in figure \ref{fig:Sync_spec}b,
i.e., $\alpha_1 = -0.67$ and $S_\nu = 141.1$ Jy 
at the frequency of $\nu =1.6$ GHz, respectively.
The synchrotron spectrum is integrated down to the frequency 
corresponding to the minimum electron Lorentz factor of 
$\gamma_{\rm min} = 100$ \citep{brunetti2001,hardcastle2011}.
The break frequency is substituted for the maximum synchrotron frequency
(i.e., $\nu_{\rm max} = \nu_{\rm b} = 218$ GHz),
because the minimum-energy field is insensitive to the maximum frequency.
For a relativistic plasma with an electron-proton energy equipartition 
(corresponding to the proton-to-electron energy-density ratio of 
$k=1$ in \cite{miley1980}),
the minimum-energy field is estimated as $B_{\rm me} = 25$ $\mu$G.
A pure electron-positron plasma ($k=0$) yields $B_{\rm me} = 21$ $\mu$G.
Such a relatively low $k$ value is typically suggested for 
jets hosted by the FR-II radio galaxies (e.g., \cite{kino2012,kawakatsu2016})

Figure \ref{fig:Mag} points out that 
the magnetic field obtained from the cooling break
is stronger than that under the minimum-energy condition
by a factor of $B_1/B_{\rm me}\sim 4$--$10$ (for $k=1$). 
Even if the possible orientation of the lobe ($\theta \gtrsim 20^\circ$) 
is considered, 
the corresponding magnetic field is suggested to remain higher 
than the minimum-energy value as $B_1/B_{\rm me}\gtrsim 2$. 
Because the minimum-energy condition is thought to be nearly equivalent 
to the energy equipartition between the magnetic field and particles, 
this result claims the magnetic-field dominance 
in the northern inner lobe of Centaurus A.

It is well known through IC X-ray studies that the lobes of typical FR-II radio galaxies 
tend to follow the equipartition condition \citep{kataoka2005}
or particle dominance (e.g., \cite{isobe2002, isobe2005}).
In addition, 
a significant fraction of the hot spot associated with FR-II radio galaxies 
is suggested to exhibit a magnetic field considerably weaker than 
the equipartition or minimum-energy fields \citep{hardcastle2004,kataoka2005}. 
The radio galaxy Centaurus A is normally categorized into FR-I sources.
Therefore, the magnetic dominance in its northern inner lobe
is possible to indicate some differences of the particle acceleration mode 
in the jets between the FR-I and FR-II objects. 

Alternatively, the magnetic dominance in the northern inner lobe 
of Centaurus A is potential to be ascribed to its compact nature,
and correspondingly to its youthfulness. 
The size of this lobe is found to be as small as 
those of the well-studied FR-II hot spots \citep{hardcastle2004,kataoka2006} 
as displayed in figure \ref{fig:Hillas}. 
When its longitudinal size ($L_{\rm c1} = 3.3$ kpc) 
is employed as the cooling length of the lobe,
the derived cooling break frequency gives a radiative age 
as $\sim 0.1$ Myr $(v/0.1c)^{-1}$.
This is by two orders of magnitude shorter 
than the typical age, $\sim 10$ Myr \citep{jamrozy2008,harwood2013},
of the well-studied FR-II radio galaxies.
It is anticipated that 
the jet-associated structures in the radio galaxies 
make a transition from the magnetic dominance to particle one
through their life. 

As mentioned in section \ref{sec:intro},
the IC technique is more sensitive to sources 
exhibiting a higher particle dominance.
Thus, the magnetic dominance revealed by the cooling-break method 
clearly demonstrates the advantage of this method over the IC technique. 
A systematic application of the cooling-break method to the radio galaxies 
is strongly requested to uncover the controlling parameters 
for the energetics associated with their jets.
For this purpose,
the PRobe far-Infrared Mission for Astrophysics (PRIMA; \cite{glenn2025}),
a cryogenically cooled far-infrared observatory concept 
selected as a candidate for the NASA Probe Explore mission,
is regarded as very useful.


\subsection{Possible contribution from invisible plasma elements}
\label{sec:proton_contrib}
The lobes and hot spots of FR-II radio galaxies 
are reported to rarely exhibit the magnetic field weaker than 
their minimum-energy or equipartition values 
\citep{hardcastle2004,kataoka2005}. 
Thus, in the following, 
possible scientific ideas are discussed to get rid of the disagreement  
between the minimum-energy condition and the cooling break 
observed from the northern inner lobe of Centaurus A. 
Because the cooling-break technique is regarded as very simple 
and physically reliable \citep{isobe2025}, 
the above minimum-energy magnetic field is assumed to be underestimated.

It is known that there are several invisible plasma elements,
which possibly enhance the minimum-energy magnetic field \citep{miley1980}.
The contribution from low-energy electrons are considered 
as the first candidate,
although it has long been under debate whether the lobes widely contains 
the electrons with a Lorentz factor of 
$\gamma_{\rm e} \ll 100$ (e.g., \cite{brunetti2001}).
By lowering the minimum electron Lorentz factor 
down to $\gamma_{\rm min} = 1$, 
the minimum-energy magnetic field was recalculated 
as $B_{\rm me} \sim 40$ $\mu$G for $k = 1$.
Thus, the minimum-energy magnetic field is found to become 
comparable to the result from the cooling break 
for only a limited condition around $\theta = 20^\circ$ and $v=0.1 c$
(see figure \ref{fig:Mag}). 

Alternatively, the energy density of protons (or heavy nuclei) is 
supposed to be fairly higher than those of electrons and magnetic fields 
(i.e. $k\gg0$).
In principle, if the protons are extremely energized through the shock,
their energy contribution is possible to reach $k \simeq 2000$ \citep{pacholczyk1970}.
Actually in the case of the northern inner lobe of Centaurus A,
the proton energy fraction of $k\gtrsim300$ yields 
the minimum-energy magnetic field as $B_{\rm me} \gtrsim 100$ $\mu$G
for $\gamma_{\rm min}=100$.
This $B_{\rm me}$ value is found to 
agree with the magnetic field corresponding to the cooling break.
The similar high-$k$ scenario was invoked 
to investigate the mid-to-far infrared excess 
observed from the west hot spot of Pictor A \citep{isobe2023,isobe2025}.
However, as mentioned above (in subsection \ref{sec:Bme}), 
there has yet been only limited concrete evidence 
for such a high proton predominance \citep{kino2012,kawakatsu2016}.
A systematic search for high $B/B_{\rm me}$ objects, 
through the cooling-break technique in particular, 
and detailed investigation to their proprieties are strongly desirable.

\subsection{Jet-to-lobe magnetic-field amplification}
\label{sec:B-amplification}
From the region around the nucleus of Centaurus A,
a detection is reported 
of diffuse VHE gamma-ray emission\citep{aharonian2009}.
Base on its spacial scale of $\sim 2.5'$
($\sim 2.4$ kpc at the distance of Centaurus A) and directional alignment 
revealed in the recent VHE study \citep{hess2020}, 
this diffuse gamma-ray emission was ascribed to the inner jet. 
By modeling the observed spectral energy distribution 
from the radio to VHE gamma-ray frequency ranges 
by the synchrotron and IC components, 
they roughly estimated the magnetic field in the inner jet as $B=23$ $\mu$G. 
However, they noted that the adopted radio-to-infrared spectrum 
should be treated as an upper limit.
This means that the derived magnetic field is 
also regarded as an upper limit. 
The magnetic field in the northern inner lobe 
measured by the cooling break approach
is found to be larger than that of the inner jet 
by a factor of $\sim 4$ (even $\sim 2$ for $\theta = 20^\circ$).
Thus, some field magnification processes are inferred  
to be operated between the jet and lobe.

Analytical (e.g., \cite{fraschetti2013}) and 
magnetohydrodynamic (e.g., \cite{inoue2009,sano2012,ji2016}) 
studies show that 
non-relativistic strong shock is possible to boost  
the downstream magnetic field by up to two orders of magnitude,
through magnetic turbulence 
caused by plasma instabilities and inhomogeneities.
Similarly, relativistic shocks are theoretically predicted to strengthen   
the post-shock magnetic field, e.g., by more than a factor of 10 
(\cite{inoue2011,mizuno2011}), via the turbulence.
These mechanisms are successfully applied to 
X-ray shells, rims and hot spots,
which exhibit a relatively strong magnetic field
\citep{bamba2003,vink2003,uchiyama2007},
hosted by Galactic supernovae remnants.
A similar scenario was employed to interpret the high $B/B_{\rm me}$ ratio 
suggested for the mid-to-far infrared excess 
in the western hot spot of Pictor A \citep{isobe2020,isobe2023}.
The magnetic-field amplification due to the post-shock turbulence 
is also regarded as very attractive to interpret 
the strong magnetic field in the northern inner lobe of Centaurus A.
However, 
it is unclear that the magnetic-field enhancement is 
fully compatible to previous observational data of this object,
while the synchrotron surface brightness in the radio and mid-infrared bands
is reported to be by a factor of $\sim3$ strengthened through the flare point.

\citet{hardcastle2011} investigated simultaneously 
the spatial distribution of the radio synchrotron emission 
from the inner jet of Centaurus A 
and the multi-frequency spectral energy distribution 
between the radio and VHE gamma-ray bands,
by assuming that the VHE gamma-ray emission from Centaurus A 
is ascribed to the IC emission from the diffuse component 
inside the inner jet \citep{kraft2002,kataoka2006}. 
Through a detailed synchrotron and IC calculation, 
which incorporates all the relevant physical processes 
and all the related IC seed photon sources, 
they indicated the magnetic field in the inner jet 
is comparable to or possibly higher 
than the equipartition field strength. 
A combination of the present study and 
that of \citet{hardcastle2011} 
maybe proposes an interesting picture that
the magnetic-field dominance is a common property
between the inner jet and inner lobes of Centaurus A.

It is strongly requested to observationally investigate 
the role of magnetic field, especially of its turbulence, 
in the northern inner lobe of Centaurus A.
For this purpose,

it is of prime importance to polarimetrically observe the inner lobe,
especially the region around the flare point, 
with the Atacama Large Millimeter/submillimeter Array (ALMA)
in a sub-arcsec spatial resolution.

\subsection{Centaurus A as a cosmic-ray accelerator}
\label{sec:CR_source}
Finally, the scientific influence  of the present result 
is briefly discussed from the viewpoint 
of the origin of the UHECRs with an energy of $> 10^{18}$ eV.
Sky regions with an excess UHECR flux in an angular scale of a few $10^\circ$ 
are widely recognized through recent UHECR observations (e.g., \cite{aab2018,tkachev2021}).
Because the UHECR flux is predicted to be strongly attenuated 
by interactions with the cosmic background radiation 
(the so-called GZK effect; \cite{greisen1966,zatsepin1966}),
their sources are obliged to be located within the distance of 
at most $\lesssim 100$ Mpc (e.g., \cite{harai2006,kotera2011}).
Thus, the above anisotropy is inevitably attributed to nearby objects,
including starburst galaxies \citep{aab2018,abdul2024}, 
radio galaxies \citep{matthews2018}, and so forth. 
Thanks to its proximity as the nearest known active galaxy,
Centaurus A is frequently mentioned as the major contributor to the UHECR anisotropy 
(e.g.,
\cite{romero1996,abraham2007,moskalenko2009,kim2013,
    matthews2018,mollerach2022,mollerach2024}.)

\begin{figure}[t]
    \centering
    \includegraphics[width=0.95\linewidth]{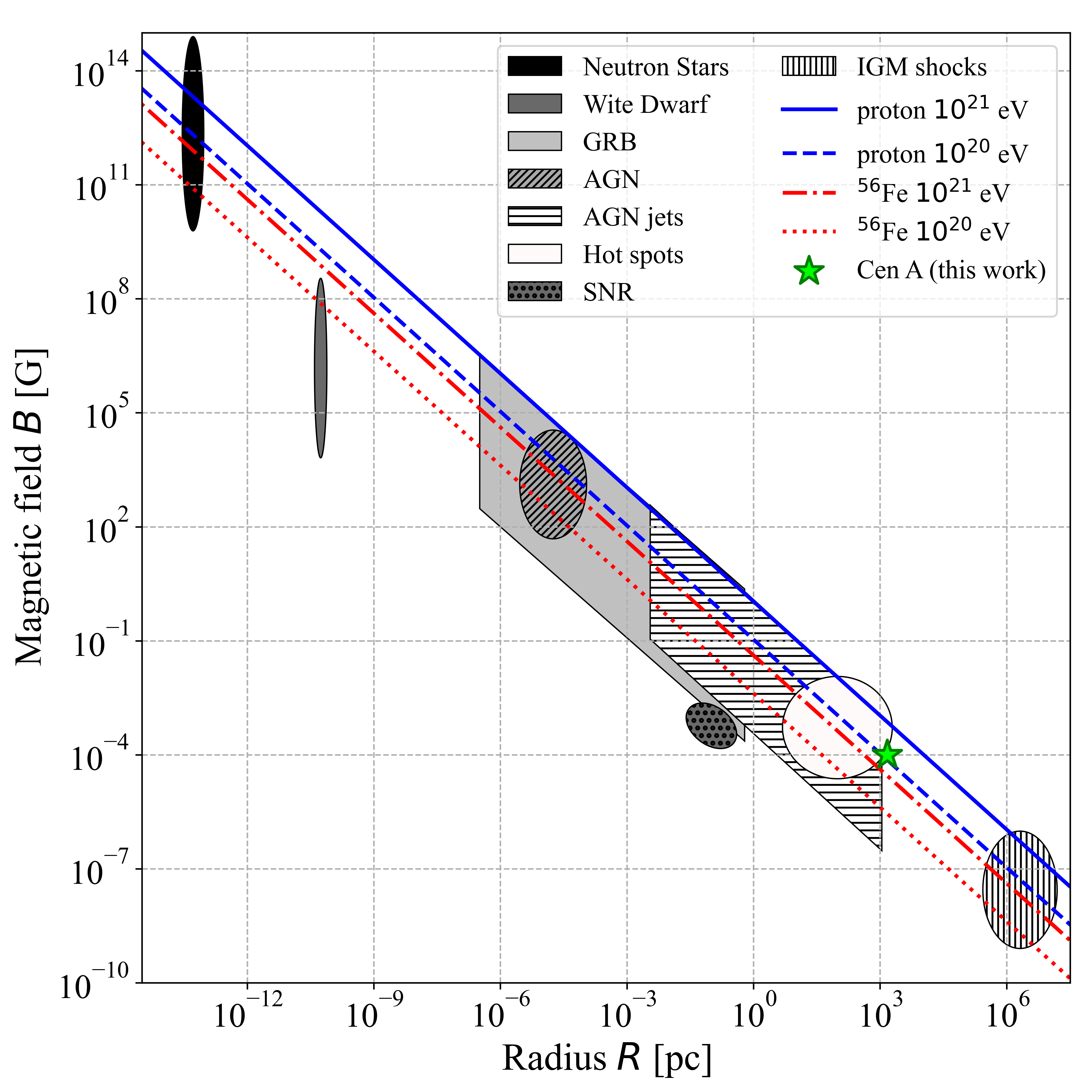}
    \caption{
    Relation between the radius $R$ and magnetic field $B$ of 
    various classes of celestial objects 
    (so-called the Hillas diagram; \cite{hillas1984}).
    The typical $B$ and $R$ values 
    of possible astrophysical UHECR accelerators 
    are taken from \citet{kotera2011}.
    The northern inner lobe of Centaurus A is indicated with the star. 
    The proton Larmor radius is drawn with the solid and dashed line, respectively,  
    for the energy of $10^{21}$ and $10^{20}$ eV. 
    The dash-dotted and dotted lines, respectively,  
    show the Larmor radius of $10^{21}$ and $10^{20}$ eV iron 
    ($^{56}$Fe) nuclei. \\
    }
    \label{fig:Hillas}
\end{figure}

Figure \ref{fig:Hillas} displays 
the relation between the typical size (radius) $R$ and magnetic field $B$ 
of various astrophysical particle accelerators, 
taken from \citet{kotera2011}.
The figure, usually called the Hillas diagram \citep{hillas1984},
is widely utilized to judge upto what energy 
a celestial object is possible to confine accelerated particles. 
The solid and dashed lines, respectively, in figure \ref{fig:Hillas}
indicate the Larmor radius of protons 
with an energy of $10^{21}$ and $10^{20}$ eV.
Recent observations suggest that 
a significant fraction of the UHECRs is 
composed of heavy nuclei \citep{kotera2011,abbasi2024,abbasi2025}.
Thus, the Larmor radius of an iron nucleus ($^{56}$Fe) 
is also drawn with the dash-dotted and dotted lines 
for the energies of $10^{21}$ and $10^{20}$ eV, respectively,
since the iron nucleus is inferred to exhibit 
a relatively large attenuation length,
which is comparable to that of protons \citep{harai2006}.

In figure \ref{fig:Hillas},
the northern inner lobe of Centaurus A is plotted with the star 
at the position of $R=1.5$ kpc and $B_1\simeq100$ $\mu$G,
the former of which corresponds to the radius along the minor axis 
(see section \ref{sec:Bme}).
The size of this lobe is larger than the Larmor radius of $10^{20}$ eV protons 
(the dashed line in figure \ref{fig:Hillas}) and 
that of $10^{21}$ eV iron nuclei (the dash-dotted line) for the derived magnetic field. 
Potentially, the protons and iron nuclei with an energy 
of $1.4 \times 10^{20}$ eV and $3.6 \times 10^{21}$ eV 
are confined inside the object.
This result has significantly enhanced the importance of Centaurus A 
as a generator of the highest-energy UHECRs. 
It is strongly recommended to investigate the particle acceleration mechanism 
in detail in the northern lobe of Centaurus A, 
although it is out of the scope of the present paper.

\begin{ack}
The authors appreciate the constructive comments 
from the anonymous reviewer to improve the present paper.
The authors are grateful to Dr. M. Hardcastle 
for his kindly providing the published $1.6$ and $4.9$ GHz 
radio images of Centaurus A in the FITS format. 
This research has been made use of the FIR data 
taken with the Herschel space observatory,
an ESA space observatory with science instruments 
provided by European-led Principal Investigator consortia and 
with important participation from NASA.
\end{ack}

\section*{Funding}
\label{sec:funding}
The authors were supported by JSPS KAKENHI Grant Number 
21H04496, 21K03635, 22H00157, 23H00130, 23H00134, 23H05441, 
23K17695, and 25K24561. 

\section*{Data availability} 
\label{sec:data_availability}
The SPIRE and PACS data utilized in the study are electrically available 
at the Herschel Science Archive \footnote{{\tt https://archives.esac.esa.int/hsa/whsa/}}.
The $1.4$ GHz radio date of Centaurus A was taken from the NASA/IPAC Extragalactic Database 
\footnote{{\tt https://ned.ipac.caltech.edu/}}.
The $1.6$ and $4.9$ GHz radio data in the electric form, 
though publicly unavailable, 
were given by Dr. M. Hardcastle through private communication.




\bibliographystyle{aasjournal}
\bibliography{main.bib}


\end{document}